\documentclass[preprint,
amsmath,amssymb, aps,
]{revtex4}

\newcommand{\be}{\begin{equation}}
\newcommand{\ee}{\end{equation}} 
\newcommand{\bea}{\begin{eqnarray}} 
\usepackage{graphicx}
\newcommand{\eea}{\end{eqnarray}}
\newcommand{\bs}{\boldsymbol}
\usepackage{epsfig}
\usepackage{bm} 

\usepackage[dvipsnames]{xcolor} 

\begin{document}

\title{Hydrodynamic Decay of Decorated Quantum Vortex Rings}
\author{L. Moriconi}
\affiliation{Instituto de F\'\i sica, Universidade Federal do Rio de Janeiro, \\
C.P. 68528, 21945-970, Rio de Janeiro, RJ, Brazil}

\begin{abstract}
The decay of quantum vortex rings in counterflow regimes, visualized in Helium II with the help of solid hydrogen particles trapped to their cores, has been a puzzling issue within the usual description of superfluid vortex dynamics, grounded on the hypothesis that a vortex filament is, effectively, an extended massless object subject to a canceling superposition of Magnus and mutual friction forces. We discuss, from a general energy-budget point of view, a phenomenological solution of this problem, which relies on viscous and quantum dissipation mechanisms, the later associated to the backreaction of vortex singular structures on the surrounding two-component fluid mixture.
\end{abstract}

\maketitle
\section{Introduction}

Superfluid (quantum) vortices, conjectured by Onsager in 1949 \cite{onsager} and further discussed by Feynman a few years later \cite{feynman}, 
are well-defined flow structures, which have atomic-sized singular cores, and 
dynamical evolution ruled, in the zero temperature limit, 
by the classical Euler equation \cite{donnelly}. Due to macroscopic quantum coherence,
the circulation around superfluid vortices is quantized in integer multiples of $h/m$ ($h$ 
is the Planck constant and $m$ is the mass of the superfluid particle constituents). These topologically 
stable excitations are assumed to play fundamental roles in the description of both equilibrium and 
out-of-equilibrium superfluid phenomena. Actually, while the 
suppression of \textcolor{black}{superfluid phase coherence} near the critical temperature
is related to the proliferation of quantum vortex loops 
induced by thermal fluctuations \cite{williams,shenoy}, turbulence below the lambda point has been generally 
depicted as a complex tangle of superfluid vortices \cite{halp_tsu,cooper_etal}. 
A proper modeling of quantum vortex evolution is thus believed to be the key for a
phenomenological understanding of quantum turbulent regimes.

Even though objectively defined from the topological properties of the condensate wavefunction, 
and detected long ago by means of ingenious mechanical experiments \cite{vinen58,whit_zimm}, 
the direct observation of superfluid vortices has been a challenging task.
The breakthrough experiments of Yarmchuk, Gordon and Packard \cite{YGP} provided clear 
visualizations of stationary arrays of He-II quantum vortices in a rotating tank. However, 
their observation technique, based on electric discharges produced from electron bubbles 
trapped in the vortex cores \cite{williams_packard}, where pressure is depleted, is not suitable for the investigation of evolving vortex filaments. 

After a time gap of almost three decades, a turning point in the problem of quantum 
vortex visualization was reached through the use of micro-sized solid hydrogen or deuterium 
particles \cite{bewley_etal,chago_sci,guoa}, which are often trapped in quantum vortex cores. 
This approach has paved the way for an ongoing wave of new experiments, 
as the visualization of vortices near the superfluid transition \cite{bewley_etal}, 
the study of quantum vortex reconnection \cite{bewley_etal2}, the decay of quantum vortex rings \cite{bewley_sree} 
-- our main focus in this work -- and the observation of Kelvin waves in vortex filaments \cite{fonda_etal}.
Interesting open challenges are furthermore related to the amount of dynamic information that one can
obtain from the recorded trajectories of tracking particles in superfluid turbulent regimes \cite{zhang_sci,serg_bar_kivo,kivo_wilkin,duda,mastracci_guo,mori,guo_etal}.

\textcolor{black}{The visualization of particle loaded vortices yields, furthermore, an interesting
stage for the validation (or not) of the vortex line structural approach to quantum vortex dynamics 
pioneered by Schwarz \cite{schwarz}, which stands at a central place in the literature}.
A careful experiment showed, \textcolor{black}{in this context}, that the application of the structural 
modeling framework to the decay of a quantum vortex ring decorated by attached solid microscopic particles 
is plagued with difficulties \cite{bewley_sree}. In that study, an estimate of the vortex ring decay 
time scale was attempted, under the assumptions that 
\vspace{0.2cm}

\noindent (i) the vortex ring core size should 
be modified by the presence of the attached particles, 
\vspace{0.2cm}

\noindent (ii) the vortex filament should experience an additional drag force due to the 
presence of normal flow around the attached particles, and that 
\vspace{0,2cm}

\noindent (iii) the existing background counterflow should 
not affect the vortex ring evolution. 
\vspace{0,2cm}

\noindent While assumption (i) is perfectly reasonable, (ii) and (iii) 
are {\it{ad hoc}} hypotheses of difficult justification. Also, it turns out that even along these 
lines the predicted time evolution of the vortex ring radius does not accurately match observations.

A main point of criticism to the usual vortex line structural description is that it completely 
neglects backreaction effects of quantum vortex filaments on the surrounding normal component of the
condensate. They are particularly relevant in situations where vortex reconnection events are assumed 
to play a significant dynamical role \cite{barenghi_etalbook,villois_etal}. One
expects, actually, normal flow to be always induced by quantum vortex filaments. 
As a surprising illustration of this fact, it has been suggested that superfluid vortex rings exhibit 
an interesting triple comoving flow structure composed of two vortex rings of normal fluid coupled to 
the much thinner quantum vortex ring \cite{kivotides_etal,galan_etal}. Our aim in this work is to show 
that similar backreaction effects offer a phenomenologically consistent and accurate solution of the 
vortex ring decay puzzle. Our discussion is based on a general energy-budget perspective, which 
bypasses specific modeling details and places emphasis on the role of normal viscous dissipation 
and normal/superfluid mutual interaction effects.

This paper is organized as follows. In Sec. II, we briefly review the 
basic ideas of the vortex line structural approach and point out modeling 
difficulties related to the phenomenon of particle-decorated vortex ring decay. 
In sec. III, we address a general phenomenological discussion 
of the vortex ring evolution along energy budget lines, which takes into account 
the coupling between the superfluid and the normal components of the flow. We devise,
in this way, a time evolution equation for the vortex ring radius which leads to a fine agreement 
with experimental data. Finally, we summarize, in Sec. IV, our results and indicate
directions of further research.


\section{Structural Modeling Issues}

The motion of a quantum vortex filament in a superfluid condensate which has
assigned velocity fields ${\bf{ v_n}}({\bf{ r}},t)$ and ${\bf{ v_s}}({\bf{ r}},t)$  for its normal and superfluid
components, respectively, is subject to Magnus and drag forces \cite{donnelly,schwarz}. Thus, if a point
$P$ of the vortex filament where the local unit tangent vector is $  {\bs{\hat \omega}}_{\bf{s}} $ (oriented along
the vorticity field) moves with velocity ${\bf{v_L}}$, the Magnus and drag forces
forces per unit length at $P$ are, respectively,
\bea
\hspace{-0.8cm} {\bf{F_M}} &=& \rho_s \kappa {\boldsymbol{\hat \omega}}_{\bf{s}} \times ( {\bf{v_L}} - {\bf{v_s}}) \ , \ \label{FM} \\
\hspace{-0.8cm} {\bf{F_D}} &=&  D_L {\boldsymbol{\hat \omega}}_{\bf{s}} \times ({\bf{v_n}} - {\bf{v_L}}) +
 D {\boldsymbol{\hat \omega}}_{\bf{s}} \times \left [ {\boldsymbol{\hat \omega}}_{\bf{s}} \times ({\bf{v_L}} - {\bf{v_n}}) \right ] \ , \ \label{FD}
\eea
where $\rho_s$ is the superfluid density, \hbox{$\kappa \simeq 10^{-7} $ m$^2$/s} is the quantum of circulation, and $D_L$ and $D$ are phenomenological drag coefficients that model the interaction between the normal and superfluid 
components of the condensate. \textcolor{black}{Assuming} that a quantum vortex filament has negligible inertia, we obtain
\be
{\bf{F_M}} + {\bf{F_D}} = 0 \ , \  \label{FMFD}
\ee
which in view of Eqs. (\ref{FM}) and (\ref{FD}) gives, for the vortex filament velocity \cite{schwarz},
\be
{\bf{v_L}} = {\bf{v_s}} + \alpha  {\boldsymbol{\hat \omega}}_{\bf{s}} \times ({\bf{v_n}} - {\bf{v_s}}) - \alpha' {\boldsymbol{\hat \omega}}_{\bf{s}} \times \left [ {\boldsymbol{\hat \omega}}_{\bf{s}}\times ({\bf{v_n}}- {\bf{v_s}}) \right ]
\label{vL} \ , \ 
\ee
where
\be
\alpha \equiv \frac{ \rho_s \kappa D}{D^2 + D_0^2} \hbox{~~~and~~~} \alpha' \equiv 1 - \frac{ \rho_s \kappa D_0}{D^2 + D_0^2} \label{alphas}
\ee
are the so-called mutual friction coefficients, with
\be
D_0 \equiv \rho_s \kappa - D_L \ . \  \label{D0}
\ee
\textcolor{black}{Eq. (\ref{FMFD}) is a fundamental postulate in the structural description of the vortex line, to be regarded more as a working hypothesis, rather than a well-established truth. The conjecture that vortex filaments are massless objects has been in fact the subject of relevant questioning \cite{toikka-brand, simula}}.

The decay of a decorated, fully-visualized, quantum vortex ring with initial radius {\hbox{$R = 400$ $\mu$m}} and lifetime of about \hbox{6 s}, subject to an approximately uniform background counterflow of velocity $\simeq 500$ $\mu$m/s, was meticulously observed by Bewley and Sreenivasan \cite{bewley_sree}. The Bewley-Sreenivasan (BS) experiment was performed at the temperature \hbox{$T=2.06$ K}, where the He-II mutual friction coefficients are \hbox{$\alpha \simeq 0.37$ and $\alpha' \simeq 3 \times 10^{-3}$} \cite{bar-donn}. 

Once Eqs. (\ref{vL}-\ref{D0}) are supposed to describe the evolution of vortex filaments, it is natural to presume, as a first modeling attempt, that the decay of decorated vortex rings could be ruled by similar equations, with appropriate redefinitions of the drag coefficient parameters and a reasonable choice of the counterflow velocity field (which may be difficult to measure in practice). This is the point of view taken in Ref. \cite{bewley_sree}, which we momentarily adopt for the sake of critical analysis.

In the particular situation of the BS experiment, the decaying vortex ring was not seen to rotate or to have its circular shape deformed during its evolution. It follows, then, as a consequence of Eq. (\ref{vL}), that ${\bf{v_{ns}}} =  {\bf{ v_n}} - {\bf{v_s}}$ should be a vector field normal to the vortex ring plane. 
Denoting by 
${\bf{\hat n}}$ a unit vector which is parallel to the self-induced vortex ring translation velocity, we may write
\be
{\bf{v_{ns}}} \equiv v_{ns} {\bf{\hat n}} = (U_{ns} - v^\circ_s ) {\bf{\hat n}} \ , \ 
\ee
where $U_{ns}$ is the approximately uniform background counterflow velocity and $v^\circ_s > 0$, with $v^\circ_s \ll |U_{ns}|$, is the self-induced vortex ring velocity. Taking into account that $\alpha' \ll 1$ and that ${\bf{v_s}}$ is also approximately uniform on the vortex ring filament, it is not difficult to get, from Eq. (\ref{vL}), the evolution equation for the vortex ring radius $R$,
\be
\frac{d R}{dt} = \alpha ( U_{ns} - v^\circ_s ) \ . \ \label{Rdot1}
\ee
Quantum vortex rings have hollow cores, where the superfluid condensate vanishes. For a core of radius \hbox{$r_c \ll R$}, the self-induced
vortex ring translation velocity is given by \cite{saffman}
\be
v_s^\circ = \frac{\kappa}{4 \pi R} \left [  \ln \left ( \frac{8R}{r_c} \right ) - \frac{1}{2}  \right ] \ . \  \label{vs_ring}
\ee
The vortex ring core radius $r_c$ depends on the temperature. For temperatures close to $2$ K, as reported in the BS experiment, we estimate $r_c \simeq 10$ $\mathring{\mathrm{A}}$ \cite{hall}.

Before proceeding with the application of (\ref{Rdot1}) and (\ref{vs_ring}) to the problem of decorated vortex ring decay, it is important to emphasize a couple of related phenomenological points:
\vspace{0.2cm}

\noindent (i) Particles that are trapped to quantum vortices are constrained to passively follow their host vortex filaments. In fact, bouyant/pressure forces or even forces associated to the normal component of the condensate, like the viscous drag, are negligible if compared to the strong restoring superfluid pressure forces that act on the particles near vortex cores.
\vspace{0.2cm}

\noindent (ii) Vortex filaments can be populated by a largely variable number of attached particles. We can distinguish two limiting cases in our analysis, related to sparse or dense vortex ring decorations, as it is illustrated in {\hbox{Fig. 1}}. Theses decoration regimes can be parametrized by the packing parameter 
\be
p \equiv \frac{N r_p}{\pi R} \ , \ 
\ee
where $N$ is the number of attached particles and $r_p$ is their mean radius. 
\vspace{0.2cm}

We examine, in the following, what the vortex line structural description would tell us about vortex ring decays for the $p \ll 1$ and $p \gg 1$ cases.
\vspace{0.2cm}

\begin{figure}[t]
\hspace{0.0cm} \includegraphics[width=0.6\textwidth]{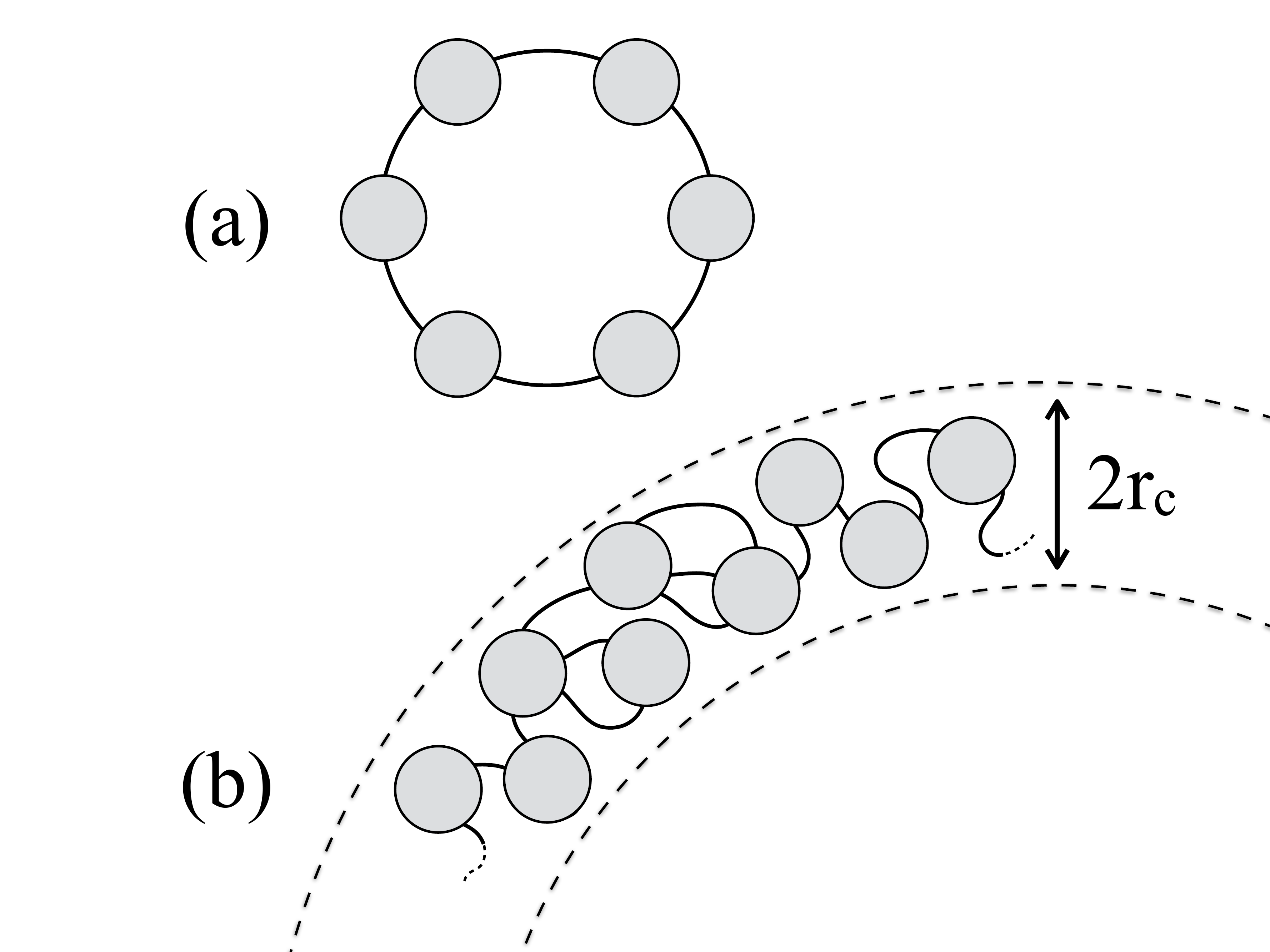}
\caption{Sparse (a) and dense (b) particle decorations of quantum vortex rings. In the packed case (b) (only a vortex ring arc is depicted), the effective vortex core radius $r_c$ can be much larger than the vortex filament core radius. Note that these pictures are merely illustrative, since particles' linear sizes are typically $10^3 - 10^4$ times larger than the ones of vortex filament cores.}
\label{vr}
\end{figure}

{\it{Sparsely Decorated Vortex Rings}} ($p \ll 1$)
\vspace{0.2cm}

When the vortex ring filament is sparsely decorated, the background and self-induced superfluid flow velocity fields are only weakly perturbed by the attached particles, so that we just have to combine Eqs. (\ref{Rdot1}) and (\ref{vs_ring}) to define and numerically solve
\be
\frac{d R}{dt} = \alpha \left \{ U_{ns} - \frac{\kappa}{4 \pi R} \left [  \ln \left ( \frac{8R}{r_c} \right ) - \frac{1}{2}  \right ] \right \} \ . \ \label{Rdot_sparse}
\ee
Results are shown in Fig. 2, for a number of counterflow velocities, together with a comparison to experimental data. As it gets clear, this modeling scenario is not satisfactory. Visual inspections suggest, as a matter of fact, that the decaying vortex ring is densely decorated, the situation we examine now.
\vspace{0.2cm}

\begin{figure}[t]
\hspace{0.0cm} \includegraphics[width=0.6\textwidth]{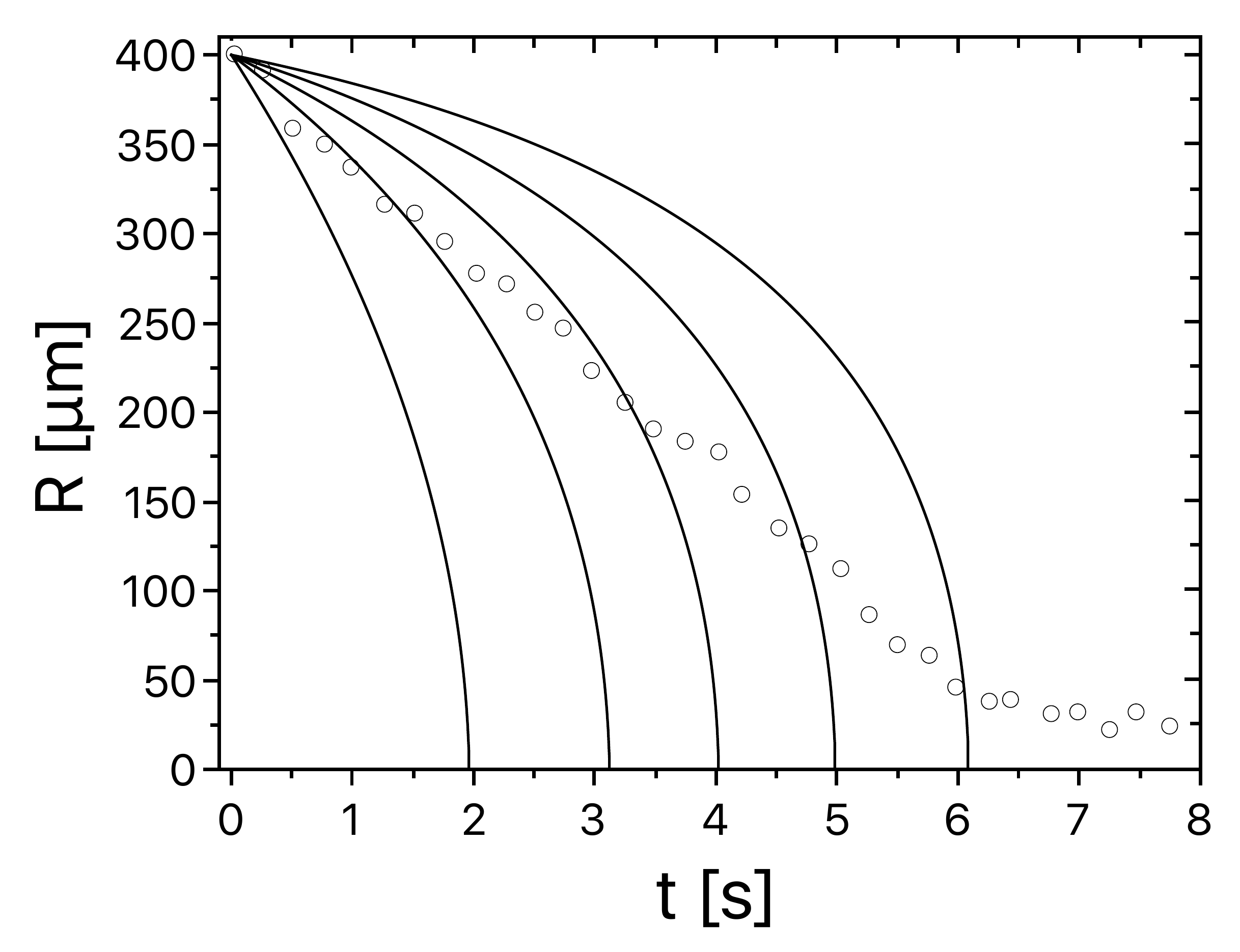}
\caption{Vortex ring radius as a function of time. Circles represent experimental data taken from Ref. \cite{bewley_sree}. Solid lines are the decaying profiles predicted from Eq. (\ref{Rdot_sparse}) for a set of counterflow velocities $U_{ns}$: in $\mu$m/s, from the left to the right in the picture,  $U_{ns} = 0$, $150$, $200$, $230$, $250$.}
\label{vr}
\end{figure}

\begin{figure}[t]
\hspace{0.0cm} \includegraphics[width=0.6\textwidth]{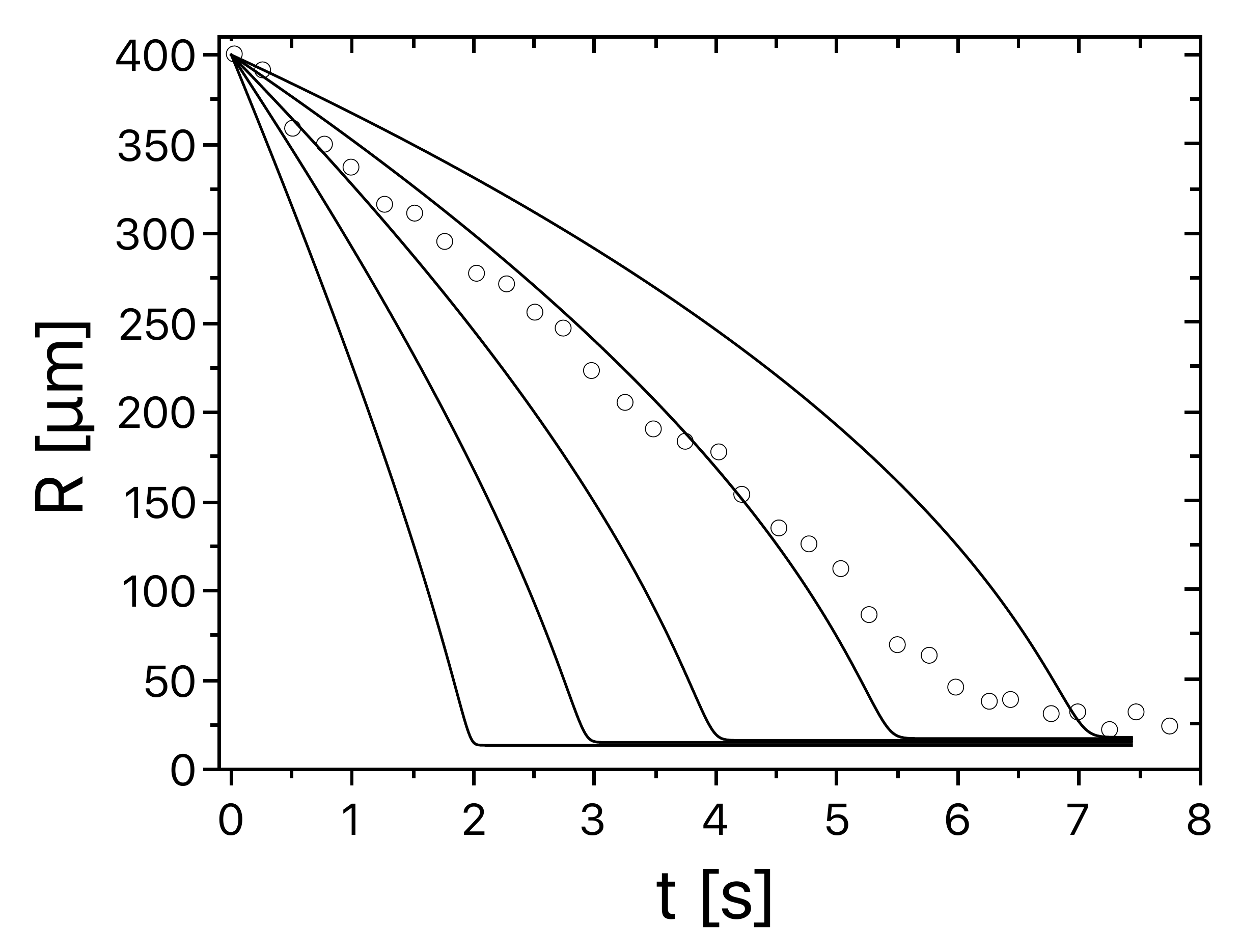}
\caption{Vortex ring radius as a function of time. Circles represent experimental data taken from Ref. \cite{bewley_sree}. Solid lines are the decaying profiles predicted from Eq. (\ref{Rdot_dense}) for a set of counterflow velocities $U_{ns}$: in $\mu$m/s, from the left to the right in the picture,  $U_{ns} = - 250$, $-120$, $-50$, $0$, $30$.}
\label{vr}
\end{figure}

{\it{Densely Decorated Vortex Rings}}  ($p \gg 1$)
\vspace{0.2cm}

A densely packed group of particles provides some non-negligible volume $V_\circ$ to the vortex ring core, which is conserved along its decay. In this case, the vortex ring radius and its core radius are related through 
\be
r_c^2 = \frac{V_\circ}{2 \pi^2 R} \ . \
\ee
This geometrical constraint implies that 
\be
\frac{R}{r_c} = c R^{\frac{3}{2}} \ , \ \label{Rrc}
\ee
where
$ c = \sqrt{2 \pi^2 / V_\circ} \simeq \pi \times 10^{-3} \mu$m$^{-\frac{3}{2}}$, as roughly estimated from snapshots of the decaying vortex ring \cite{bewley_sree}. 

It was hypothesized in Ref. \cite{bewley_sree} that the normal low Reynolds number flow around the observed densely decorated vortex ring could modify the drag coefficient parameter $D$ in (\ref{alphas}), rescaling the mutual friction coefficient $\alpha$ to the effective larger value $\tilde \alpha = 1.3 \alpha$. Replacing in Eq. (\ref{Rdot_sparse}) $\alpha$ by $\tilde \alpha$ and $R/r_c$ by $cR^{3/2}$, as prescribed in (\ref{Rrc}), we obtain
\be
\frac{d R}{dt} = \tilde \alpha \left \{ U_{ns} - \frac{\kappa}{4 \pi R} \left [  \ln \left ( 8cR^{\frac{3}{2}} \right ) 
- \frac{1}{2}  \right ] \right \} \ . \ \label{Rdot_dense}
\ee
Solutions of (\ref{Rdot_dense}) for a set of background counterflow velocities are shown in Fig. 3. A reasonable approximation to the experimental data seems to be achieved now up to the time instant where \hbox{$R \simeq 150$ $\mu$m}. The agreement, however, is based on the assumption of vanishing background counterflow, a condition of hard phenomenological support. We have also verified that solutions with alternative values of $\tilde \alpha / \alpha$, within acceptable ranges, do not improve results in a meaningful way.
\vspace{0.2cm}

The message we take from the above modeling attempts is that the usual structural vortex line approach -- including slight formulation variations -- is not enough {\it{per se}} to reproduce the evolution of densely decorated quantum vortices. After all, this should not be a matter of great surprise. As it has been pointed out in the literature, the extension of Landau's two-fluid model to cope with the existence of superfluid circulation, associated to correlated vortex filament bundles \cite{donnelly} or even singular quantum vortices \cite{kivotides_etal,galan_etal}, implies that energy and momentum should be exchanged between the normal and superfluid components of the condensate, underlying the existence of yet unsuspected phenomena.

Given the complex (and dynamic) boundary conditions involved in the geometrical characterization of a densely packed vortex ring, direct numerical simulations of the coupled normal and superfluid flows would be extremely costly for this particular problem. We address in this work, instead, a heuristic strategy to devise the general equation, in place of (\ref{Rdot_sparse}) or (\ref{Rdot_dense}), that should rule the radius evolution of densely decorated vortex rings. The argument, carried out in the next section, is essentially based on the energy balance equation derived from a two-fluid hydrodynamic model.

\section{Energy-Budget Approach}

Let $\rho_n$ and $\mu_n$ be, respectively, the density and dynamic viscosity of the normal component of the condensate. 
We take, as a model of Helium II hydrodynamics in the isothermal approximation, the Hall-Vinen-Bekarevich-Khalatnikov (HVBK) 
coupled set of equations for the normal/superfluid mixture \cite{barenghi_etalbook,galan_etal},
\bea
&& \hspace{-0.8cm} \rho_n \left ( \partial_t {\bf{v_n}} + {\bf{v_n}}  {\bs{\cdot \nabla}} {\bf{v_n}} \right )
= -  {\bs{ \nabla}} P_n + \mu_n \nabla^2 {\bf{v_n}} + {\bf{F_{ns}}} \ , \  \label{eqvn} \\
&& \hspace{-0.8cm} \rho_s \left ( \partial_t {\bf{v_s}} + {\bf{v_s}}  {\bs{\cdot \nabla}} {\bf{v_s}} \right )
= -  {\bs{ \nabla}} P_s  - {\bf{F_{ns}}}  \ , \ \label{eqvs}
\eea
supplemented by the double incompressibility condition,
\be
{\bs{\nabla \cdot}} {\bf{v_s}} = {\bs{ \nabla \cdot}} {\bf{v_n}} = 0 \ . \ \label{div_vnvs}
\ee
Above, $P_n$ and $P_s$ are the partial pressures associated to the Helium II normal and superfluid components. 
In our specific problem, the force per unit volume, ${\bf{ F_{ns}}}$, that the 
superfluid component exerts on the normal one, is assumed to have singular support 
on the vortex filament pieces that hold the particles in the vortex ring (see {\hbox{see again Fig. 1}}).
Notice that the force density ${\bf{ F_{sn}}} = - {\bf{ F_{ns}}}$ can be readily computed from the substitution of (\ref{vL}) into (\ref{FD}),
using (\ref{alphas}) and the fact that the singular superfluid vorticity field 
$ {\bs{\nabla}} \times {\bf{v_s}} $ can be represented in a small neighborhood of the vortex filament around
its position ${\bf{r_v}}$ as
\be
{\bs{\omega}}_{\bf{s}} ({\bf{r_v}} + {\bs{\xi}} ) = \kappa {\bs{\hat \omega}}_{\bf{s}} \delta^2( {\bs{\xi}} ) \ , \
\ee
where ${\bs{\xi}} \cdot {\bs{\hat \omega}}_{\bf{s}} = 0 $. We obtain, thus,
\bea
{\bf{F_{ns}}} =  \alpha \rho_s {\bs{ \hat \omega}}_{\bf{ s}} 
\times \left [ {\bs{\omega}}_{\bf{s}} \times  \left ( {\bf{ v_n}} - {\bf{v_s}} \right ) \right ] + 
\alpha' \rho_s  {\bs{\omega}}_{\bf{s}} \times \left ( {\bf{v_n}} - {\bf{v_s}} \right ) \ . \  \label{Fns}
\eea
It is important to remark that Eq. (\ref{Fns}) holds for the vortex filament segments that bridge particles in the decorated vortex ring, as depicted in Fig. 1. These various {\it{vortex bridges}} are labelled henceforth by a given set of lines $\{ C_i \}$. Furthermore, the vorticity field lines which are spread over the relatively large particle surfaces are not supposed to provide relevant contributions to the interaction force between the normal and 
superfluid flow components.

Direct numerical solutions of Eqs. (\ref{eqvn}) and (\ref{eqvs}) with the constraints (\ref{div_vnvs}) and the definition (\ref{Fns}) are far from simple even for the case of bare vortex rings \cite{kivotides_etal,galan_etal}. Considerable further complications are expected to arise in the case of decorated vortex filaments, since one would have to worry about the implementation of dynamic boundary conditions and supplementary prescriptions related to vortex reconnection events. 

It is interesting, therefore, to see how far we can proceed, having focus on general properties of the dynamical equations (\ref{eqvn}) and (\ref{eqvs}) and resorting as much as possible to a minimum number of phenomenological assumptions. To start, we write, inspired by the Reynolds decomposition procedure \cite{pope}, the normal and superfluid velocity fields as 
\be
{\bf{v_n}} = {\bf{U_n}} + {\bf{u_n}} \ , \ {\bf{v_s}} = {\bf{U_s}} + {\bf{ u_s}} \ , \  \label{us}
\ee
where 
\be
{\bs{\nabla}} \cdot {\bf{ u_s}} = {\bs{\nabla}} \cdot {\bf{u_n}} = 0 \ . \
\ee
The velocity fields ${\bf{U_n}}$ and ${\bf{U_s}}$, assumed to be time-independent and uniform, represent the background counterflow in the absence of quantum vortex filaments. The fields ${\bf{u_n}}$ and $ {\bf{u_s}}$ are, on their turn, the velocity perturbations associated to the dynamic evolution of the decorated vortex ring. We, now, 
\vspace{0.2cm}

\noindent (i) substitute (\ref{us}) in Eqs. (\ref{eqvn}) and (\ref{eqvs}) and contract them with ${\bf{u_n}}$ and $ {\bf{u_s}}$, respectively; 
\vspace{0.2cm}

\noindent (ii) sum up the contracted equations and integrate the resulting scalar equation over all the space,  taking (\ref{div_vnvs}) into account. 
\vspace{0.2cm}

\noindent In this way, we find, up to second order in ${\bf{u_n}}$ and ${\bf{u_s}}$, the energy-budget equation,
\bea
&&\frac{d}{dt} \left ( E_s + E_n \right ) = \nonumber \\
&=&  -\mu_n \int d^3 {\bf{r}} {\hbox{Tr}} \left [ 
\left ( {\bs{\nabla}} \otimes {\bf{u_n}} \right )^2 \right ] - 
\alpha \rho_s \kappa \sum_i \int_{C_i} ds  ({\bf{u_n}}- {\bf{u_s}})_\perp^2 + \nonumber \\
&-& 
\alpha \rho_s \kappa \sum_i \int_{C_i} ds  ({\bf{u_n}}- {\bf{u_s}})_\perp \cdot 
({\bf{U_n}}- {\bf{U_s}})_\perp + \nonumber \\
&+& \alpha' \rho_s \kappa \sum_i \int_{C_i} ds 
({\bf{u_n}} - {\bf{u_s}}) \cdot [{\bs{\hat \omega}}_{\bf{s}} 
\times ({\bf{U_n}} - {\bf{U_s}})]
\ , \ \label{dotEtot}
\eea
where the notation $\perp$ indicates the vector component which is normal to ${\bs{\hat \omega}}_{\bf{s}}$ and
\bea
&&E_s = \frac{1}{2} \rho_s \int d^3 {\bf{r}} [{\bf{u_s}}({\bf{r}},t)]^2 \ , \ \label{Esdef}  \\
&&E_n = \frac{1}{2} \rho_n \int d^3 {\bf{r}} [{\bf{u_n}}({\bf{r}},t)]^2 \ , \  \label{Endef}
\eea
are the superfluid and normal kinetic energy contributions related to the perturbation fields ${\bf{u_s}}$ and ${\bf{u_n}}$.

Assuming that a quantum vortex ring of time-dependent radius $R(t)$ is coupled to vortex ring-like normal flow
structures at similar length scales \cite{kivotides_etal}, it is not difficult to show, from the fluid dynamic 
Biot-Savart law \cite{saffman}, that at large distances ($r \gg R$) velocity perturbations have the self-similar form
\bea
&&{\bf{u_s}} ({\bf{r}}, t) = \frac{\kappa}{R(t)} {\bf{u}}^{\bs{(0)}}_{\bf{s}} ( {\bf{r}}/R(t))  \ , \ \label{usfar}      \\
&&{\bf{u_n}} ({\bf{r}}, t) = \frac{\kappa}{R(t)} {\bf{u}}^{\bs{(0)}}_{\bf{n}} ( {\bf{r}}/R(t)) \ , \ \label{unfar}
\eea
where ${\bf{u}}^{\bs{(0)}}_{\bf{s}} ( {\bf{r}}/R(t))$ and ${\bf{u}}^{\bs{(0)}}_{\bf{s}} ( {\bf{r}}/R(t))$ are dimensionless fields 
that depend on structural details of the normal and superfluid vortex rings (number of structures, relative positions, etc.). 
Eqs. (\ref{usfar}) and (\ref{unfar}) can be used to derive rough dimensional estimates of the several contributions in (\ref{dotEtot}). The superfluid kinetic energy (\ref{Esdef}), for instance, would be estimated as
\be
E_s = \frac{1}{2}\rho_s \kappa^2 R \int d^3 {\bf{r}} [{\bf{u}}^{\bs{(0)}}_{\bf{s}}( {\bf{r}})]^2 \propto \rho_s \kappa^2 R \ . \
\ee
However, this estimate is found to lose accuracy for vortex rings which have their core radius $r_c$ much smaller than $R$. A better approximation is (hollow vortices) given as \cite{saffman}
\
\be
E_s = \frac{1}{2}\rho_s \kappa^2 R \left [  \ln \left ( \frac{8}{e^2} \right ) + \ln \left ( \frac{R}{r_c} \right )  \right ] \ . \ \label{vrEs}
\ee
The above $\ln(R/r_c)$ contribution comes essentially from the fast growth of the velocity field intensity and approximate axial symmetry near the cores of slender vortex ring filaments, the ones which have {\hbox{$r_c \ll R$}}. An analogous log-correction term is observed for the vortex ring translation velocity, Eq. (\ref{vs_ring}).

In order to cope with both the far and the near-core vortex ring field asymptotics, we put forward estimates for the various terms in 
(\ref{dotEtot}) which contain, besides the predictions obtained from (\ref{usfar}) and (\ref{unfar}), further log-correction contributions, 
having in mind that the only relevant length scales in the system are $R$ and $r_c$. More concretely, we write
\bea
&&\star ~~~ E_n = \rho_s \kappa^2 R \left [ 
a_1 + a_2  \ln \left ( \frac{R}{r_c} \right )  \right ] \ , \ \label{En} \\
&&\star ~~~  \mu_n \int d^3 {\bf{r}}  {\hbox{Tr}} \left [  \left ( {\bs{\nabla}} \otimes {\bf{u_n}} \right )^2 \right ] 
= \frac{\nu}{R} \rho_s \kappa^2  \left [  b_1   + b_2  \ln \left ( \frac{R}{r_c} \right )  \right ] \ , \ \label{nuE} \\
&&\star ~~~  \rho_s \kappa \sum_i \int_{C_i} ds ({\bf{u_n}}-{\bf{u_s}})_\perp^2 
= \frac{\nu}{R} \rho_s \kappa^2  \left [ c_1   + c_2  \ln \left ( \frac{R}{r_c} \right )  \right ] \ , \  \label{nuE2} \\
&&\star ~~~ \rho_s \kappa \sum_i \int_{C_i} ds \left \{ ({\bf{u_n}}- {\bf{u_s}})_\perp \cdot 
({\bf{U_n}}- {\bf{U_s}})_\perp \right \}  
= U_{ns} \rho_s \kappa^2 \left [ d_1 +  d_2 \ln \left ( \frac{R}{r_c} \right ) \right ] \ , \ \\
&&\star ~~~ \rho_s \kappa \sum_i \int_{C_i} ds 
({\bf{u_n}} - {\bf{u_s}}) \cdot [{\bs{\hat \omega}}_{\bf{s}} 
\times ({\bf{U_n}} - {\bf{U_s}})] =
U_{ns} \rho_s \kappa^2 \left [ e_1 + e_2 \ln \left ( \frac{R}{r_c} \right ) \right ] \ , \
\label{ringE} 
\eea
where  $a_1$, $a_2$, ..., $e_1$, $e_2$ are unknown dimensionless coefficients to be discussed in a moment and 
$\nu \equiv \mu_n / \rho_n$ is the kinematic viscosity of the normal fluid. The attentive reader will notice that in Eqs. (\ref{En}) and (\ref{nuE}), the estimates are expressed in terms of $\rho_s$ instead of $\rho_n$. Also, the apparently extraneous kinematic viscosity prefactor $\nu$ comes into play in Eq. (\ref{nuE2}). We note, accordingly, that all of these definitions are actually just convenient ways to establish standardized estimates, which are in good terms with dimensional analysis principles.

The profusion of arbitrary coefficients in Eqs. (\ref{En}-\ref{ringE}) could induce one to think, in a first moment, that the energy-budget line of reasoning carried out so far has reached a hopeless dead end. However, there is an interesting phenomenological solution to this issue, based on the fact that as the superfluid vortex ring collapses and its energy (\ref{vrEs}) decays, all the terms in (\ref{dotEtot}) concomitantly decay as well. It is reasonable to conjecture, thus, that all the unknown coefficients in (\ref{En}-\ref{ringE}) have their ratios $a_1/a_2$, $b_1/b_2$, ..., $e_1/e_2$ approximately {\it{locked}} to the value prescribed by the known coefficients of (\ref{vrEs}). In more expicit terms, we postulate that
\be
\frac{a_1}{a_2} \simeq \frac{b_1}{b_2} \simeq {\hbox{ ... }} \simeq \frac{e_1}{e_2} \simeq \ln \left ( \frac{8}{e^2} \right ) \ . \ \label{ratios}
\ee
Taking (\ref{En}-\ref{ringE}) and (\ref{ratios}) into account (replacing approximations by equalities), it follows, from (\ref{dotEtot}),
that
\be
\frac{d E_s}{d t} = - \gamma (R) E_s \ , \ \label{dEsdt}
\ee
where, introducing a pair of dimensionless unknowns $\beta_1$ and $\beta_2$,
\be
\gamma (R) \equiv \beta_1 \frac{U_{ns}}{R} + \beta_2 \frac{\nu}{R^2}  \ . \ \label{decayrate}
\ee
The coefficients $\beta_1$ and $\beta_2$, which are linear functions of $a_2$, $b_2$, ..., $e_2$, have now the status of the only two independent adjusting phenomenological parameters that are necessary to model the radius evolution of the decorated vortex ring decay. 

\begin{figure}[ht]
\hspace{0.0cm} \includegraphics[width=0.6\textwidth]{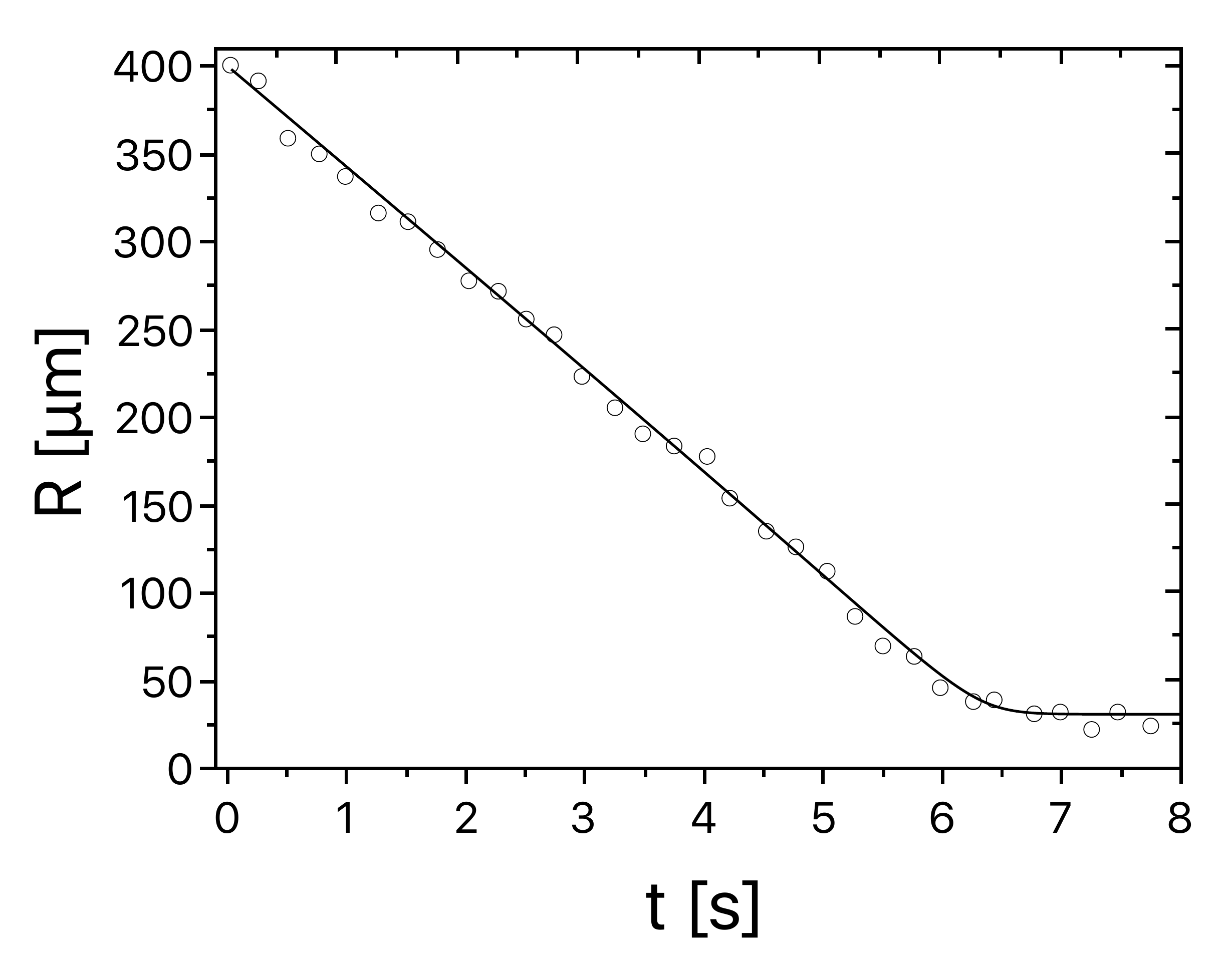}
\caption{Vortex ring radius as a function of time. Circles represent experimental data taken from Ref. \cite{bewley_sree}. The solid line is obtained from 
the numerical solution of Eq. (\ref{dRdt}) with the optimal parameters $\beta_1 = 0.14$ and $\beta_2 = 0.36$ and initial condition $R(0) = 400$ $\mu$m.}
\label{vr}
\end{figure}

The two terms in the RHS of the radius-dependent decay rate $\gamma(R)$, Eq. (\ref{decayrate}), have appealing physical interpretations: energy can be removed from the coupled normal-superfluid vortex rings by background drag/Magnus forcing (decay rates $\propto U_{ns}/R$) combined with normal viscous dissipation and excitation of normal flow near the superfluid vortex ring (decay rates $\propto \nu / R^2$). 

As already discussed in Sec. II, the vortex ring core radius is not constant, but it grows in time as the vortex ring decays,
due to volume conservation. We revisit here the estimate of the parameter $c$, introduced in Eq. (\ref{Rrc}), under the light of the present 
discussion. For this sake, we assume that the asymptotic radius $R_\infty$ that characterizes the complete collapse of the superfluid vortex 
ring is the one that leads to $E_s = 0$ in (\ref{vrEs}), that is,
\be
\frac{R_\infty}{r_c} = \frac{e^2}{8} \ . \ \label{Rrc2}
\ee
From the decay experiment data, one gets $R_\infty \simeq 30$ $\mu$m, which gives, from (\ref{Rrc}) and (\ref{Rrc2}), 
$c = 5.6 \times 10^{-3} \mu$m$^{-\frac{3}{2}}$. For the kinematic viscosity of the normal component of Helium II we take, at the temperature of $T=2.06$ K, $\nu = 1.08 \times 10^{-8} $m$^2/$s \cite{bar-donn}. The counterflow velocity is estimated as $U_{ns} = 500$ 
$\mu$m/s \cite{bewley_sree}.


Using (\ref{Rrc}) and (\ref{vrEs}), the energy dissipation equation (\ref{dEsdt}) can be reshuffled as
\bea
&&\frac{d R}{dt} = - R \gamma(R) \left [ \frac{\partial \ln(E)}{\partial \ln(R)} \right ]^{-1} \nonumber \\
&&= - \left ( \beta_1 U_{ns} + \beta_2 \frac{\nu}{R} \right ) \left [ \frac{\ln(8kR^{\frac{3}{2}}) - 2}
{\ln(8kR^{\frac{3}{2}}) - \frac{1}{2}} \right ]  \ . \ \label{dRdt}
\eea
An excellent fitting to the data (set by least squares regression) is shown in Fig. 4, produced from the numerical 
solution of (\ref{dRdt}) with $\beta_1 = 0.14$ and $\beta_2 = 0.36$. 

It is worth mentioning that direct HVBK numerical simulations of bare vortex rings 
\cite{galan_etal} were found to lead to similar linear time-dependent decaying profiles for $R(t)$. It is possible that the phenomenological 
discussion addressed in this paper can be extended to the case of decaying vortex rings which are free of attached 
particles or sparsely decorated.

\section{Conclusions}

We have investigated, within an essentially phenomenological framework, the decay of densely 
decorated quantum vortex rings. Relying upon energy balance arguments, we have been able to 
accurately reproduce the vortex ring radius measurements taken in the BS experiment \cite{bewley_sree}. 
It turns out that backreaction effects, modeled through the HVBK coupled hydrodynamical 
equations \cite{barenghi_etalbook,galan_etal}, are a fundamental ingredient to quantify the exchange of 
kinetic energy between the superfluid vortex ring and its surrounding normal (viscous) fluid. 

The assumption that quantum vortices are massless
\textcolor{black}{is by no means obvious even at zero temperature, since quantum 
excitations may play a role in processes associated to superfluid linear 
momentum variations}. As a consequence, vortex filaments may interact and evolve in completely different ways 
as the ones predicted from the Schwarz's structural approach \cite{schwarz}.

It would be interesting to investigate, under the light of the present findings, the modifications that 
should be implemented in the standard structural description of vortex dynamics, as encoded in Eq. (\ref{FMFD}), 
so as to establish alternative ways to model quantum vortex motion. One gets back, 
in this connection, to the long-time debated topic of quantum vortex inertia \cite{toikka-brand,simula}.

In order to derive the radius evolution equation (\ref{dRdt}), we had to resort to simple and reasonable
hypotheses, like (\ref{ratios}), that could be validated through further numerical simulations 
of the HVBK equations. We do not mean necessarily the numerical study of particle-laden superflows, 
which may be too heavy for the analysis of realistic cases, but simplified numerical solutions where,
as an effective boundary condition, the vortex ring core is prescribed to have a fixed volume
as the vortex ring decays.

\acknowledgements
This work has been partially supported by CNPq under grant number 307659/2015-1 .

\end{document}